%% file: SecDec2014.tex
\documentclass[a4paper]{jpconf}
\usepackage{graphicx,amsmath,amssymb}
\newcommand{\secdec}{{\textsc{SecDec}}}
\def\eps{\epsilon}
\input definitions

\usepackage{subfigure}
\graphicspath{{plots/}}

\begin{document}
\title{Numerical multi-loop calculations with the program SecDec}

\author{Sophia Borowka$^1$, Gudrun Heinrich$^{2,}$\footnote[3]{Speaker; presented at the conference ACAT 2014,
Prague, Czech Republic, September 2014.}}

\address{$^1$ Institute for Physics, University of Zurich, Winterthurerstr.190, 8057 Zurich,
Switzerland}
\address{$^2$ Max Planck Institute for Physics, F\"ohringer Ring 6, 80805 Munich, Germany}

\ead{sborowka@physik.uzh.ch, gudrun@mpp.mpg.de}

\begin{abstract}
\secdec{} is a program which can be used for the 
evaluation of parametric integrals, in particular multi-loop integrals. 
For a given set of propagators defining the graph, 
the program  constructs the graph polynomials, factorises the endpoint singularities, 
and finally produces a
Laurent series in the dimensional regularization parameter, 
whose coefficients are evaluated numerically.
In this talk we discuss various features of the program,  
which extend the range of applicability.
We also present a recent phenomenological example of an application
entering the momentum dependent two-loop corrections to neutral Higgs boson masses 
in the MSSM.
\end{abstract}

\section{Introduction}

In view of the absence of ``smoking gun" signals of new physics at the LHC so far, 
the importance of precision calculations cannot be over-emphasized. 
This means that higher oder corrections in both the QCD and the electroweak sector 
need to be evaluated, ideally without neglecting mass effects, and including 
matched parton showers as far as possible.

These calculations have many facets, however most of them have in common 
that they involve multi-dimensional integrations over some parameters, 
for example Feynman parameters in the case of (multi-)loop integrals, 
or parameters related to the integration 
over a factorized phase space of subtraction terms for infrared-divergent 
real radiation.
Usually, these calculations are performed within the framework of dimensional regularization, 
and one of the challenges is to factorise the poles in the regulator $\eps$. 

The program \secdec\,\cite{Carter:2010hi,Borowka:2012yc,Borowka:2013cma} is designed to 
perform this task, and to integrate the 
coefficients of the resulting Laurent series in $\eps$ numerically,
based on the sector decomposition algorithm described in \cite{Binoth:2000ps,Heinrich:2008si}. 
Other public implementations of sector decomposition can be 
found in \cite{Bogner:2007cr,Gluza:2010rn,Smirnov:2008py,Smirnov:2013eza}.

\section{The program \secdec}

\subsection{Basic structure}

\begin{figure}
\begin{center}
\includegraphics[width=0.9\textwidth]{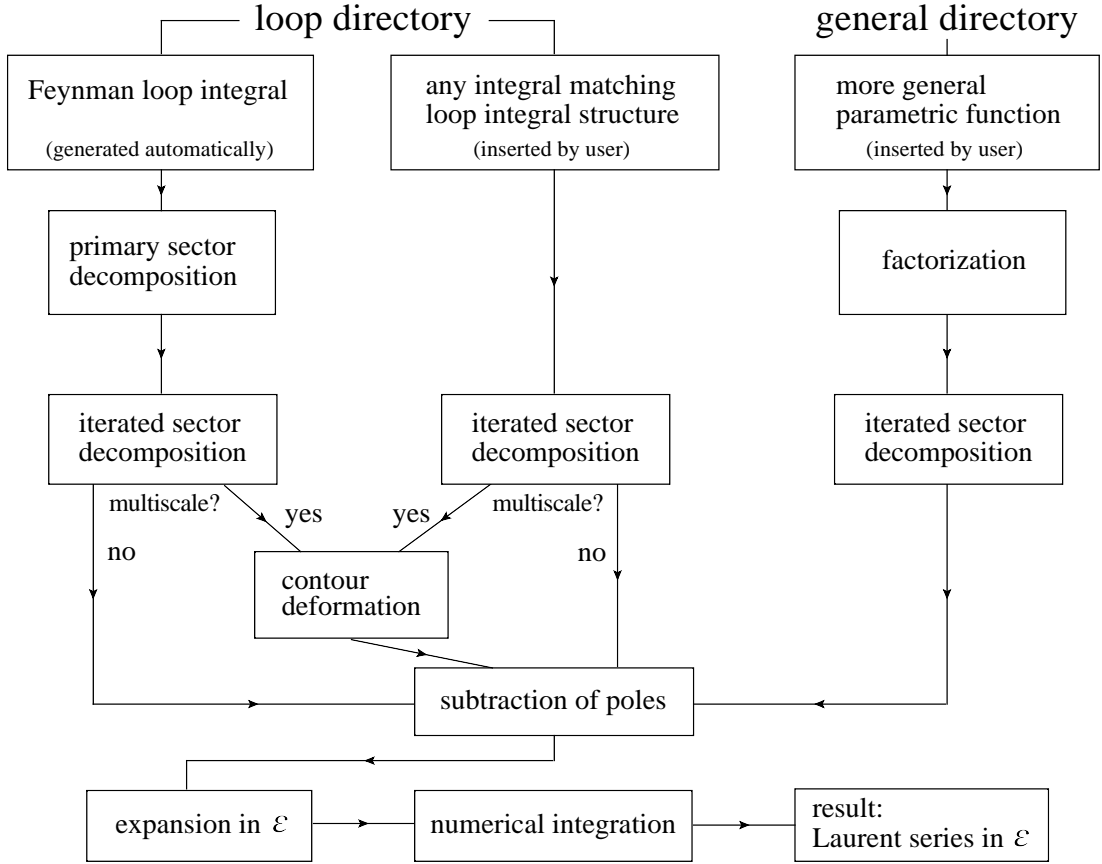}
\end{center}
\caption{Flowchart showing the main steps the program performs to produce the 
	numerical result as a Laurent series in $\epsilon$.
	\label{fig:structure} }
\end{figure}

The program consists of two main parts, one being designed for loop integrals, 
to be found in a directory called  {\tt loop}, the other one for more general parametric integrals, 
located in a directory called {\tt general}. 
The procedure to isolate the poles in the regulator $\eps$ and to do the subtractions and integrations 
is very similar in the two branches. However, only the {\tt loop} part contains the 
possibility of contour deformation, because only for loop integrals the analytic continuation 
can be performed in an automated way, following Feynman's ``$i \varepsilon$" prescription.
The basic flowchart of the program is shown in Fig.~\ref{fig:structure}.
More details about the central column will be given in section \ref{sec:userdefined}.

\subsection{Installation and usage}

\subsection*{Installation}

The program can be downloaded from 
{\tt http://secdec.hepforge.org}.
Unpacking the tar archive via 
{\it  tar xzvf SecDec.tar.gz} will create a directory called {\tt SecDec}. 
Running  {\it ./install} in the {\tt SecDec} directory will install the package.
Prerequisites are Mathematica, version 6 or above, perl 
(installed by default on most Unix/Linux systems), 
a C++ compiler, and a Fortran compiler if the Fortran option is used. 
Contour deformation is only available in C++.
The libraries {\small CUBA}\,\cite{Hahn:2004fe,Agrawal:2011tm} 
and {\small BASES}\,\cite{Kawabata:1995th}, 
which are used for the numerical integration,  
come with the package \secdec{} and will be compiled at the installation stage.

\subsection*{Usage}

The user should edit two files (where templates are provided): 
a text file where  parameters like the name of the graph, the order to be expanded in $\eps$
or numerical integration parameters are set, and a file in Mathematica syntax, 
where the graph is defined, optionally either in terms of propagators or in terms of 
lines connecting numbered vertices.
The template to define the parameters for standard loop integrals is called {\tt paramloop.input}, 
and the one in Mathematica syntax is called {\tt templateloop.m}. 
It is recommended that the user copies these template files and renames them  before editing them.
We will call the edited files {\tt myparamfile.input} and  {\tt mytemplatefile.m} in the
following.
Examples for a number of specific graphs or input functions can be found in the subdirectories 
{\tt demos} of both the {\tt loop} and the {\tt general} directory.
After having edited the input files, \secdec{} is called as follows:
\begin{enumerate}
\item from the {\tt loop} (or {\tt general}) 
directory, execute the following command in the shell:\\
{\it ./launch -p myparamfile.input -t mytemplatefile.m} 
\\
If your files {\tt myparamfile.input, mytemplatefile.m} are in a different directory, say, 
{\it myworkingdir}, 
 use the option {\it -d myworkingdir}, i.e. the full command then looks like 
 {\it ./launch -d myworkingdir -p myparamfile.input -t mytemplatefile.m}, 
 executed from the directory {\tt SecDec/loop} or
 {\tt SecDec/general}. 
 
 If  the option {\it -p myparamfile.input} is omitted, the file {\tt paramloop.input} will be taken as default.
Likewise, if the option {\it -t mytemplatefile.m} is omitted, 
the file {\tt templateloop.m} will be taken as default.
\item Collect the results. If the calculations are done sequentially on a single machine, 
    the results will be collected automatically.
If the jobs have been submitted to a cluster,    
	when all jobs have finished, use  the command 
	{\it perl results.pl [-d myworkingdir -p myparamfile]}. 
In both cases, the files containing the final results will be located in the {\tt graph} subdirectory
	specified in the input file.

\end{enumerate}

\subsection{Topology definition}

In the Mathematica input file {\tt mytemplatefile.m}, the graph can be defined in two 
ways: either by labeling the vertices and 
listing the connections between vertices (corresponding to the flag {\tt cutconstruct=1}), 
or by specifying the momentum flow explicitly (corresponding to {\tt cutconstruct=0}).  
The program will then construct the integrand in terms of 
Feynman parameters automatically.
For tensor integrals, i.e. integrals with loop momenta in the numerator, 
the propagators have to be given in terms of momenta, i.e. 
{\tt cutconstruct=0} must be used, as 
tensor integrals are not shift invariant.

An example for the construction of the graph polynomials based on labelled vertices 
is given in subsection \ref{sec:hh}.

\subsection{Controlling the different stages of the calculation}

As the program consists of a purely algebraic part and a numerical part, 
it can be useful to perform the calculation only up to a certain level, 
instead of launching immediately the full program chain including the numerical integration.
For example, to get an idea about the pole structure of an integral, 
one can first perform only the iterated sector decomposition to factor out the 
parameters exhibiting the  poles in $\eps$ to expect. This can be achieved by setting 
 {\tt exeflag=0} in {\tt myparamfile.input}. 
The following stages can be selected with the {\tt exeflag}:  
\begin{itemize}
\item 0: The iterated sector decomposition is done. The scripts to do the subtractions,  
the expansion in epsilon and to launch the numerical part 
are created (scripts batch* in the subdirectory {\it graph}), but not run. 
\item 1: The subtraction and epsilon expansion is 
performed and the resulting functions are written to Fortran/C++ files. 
\item 2: All the files needed for the numerical integration are created.
\item 3: The compilation of the Fortran/C++ files 
is launched to make the executables. 
\item 4: The executables are run to perform the numerical integration.
\end{itemize}

\subsection{Evaluation of user-defined functions with arbitrary kinematics}
\label{sec:userdefined}

In the standard setup, the program will construct the graph polynomials 
$\mathcal{F}$ and $\mathcal{U}$ and then proceed directly to the 
so-called {\it primary sector decomposition}\,\cite{Binoth:2000ps,Heinrich:2008si}, 
which serves to integrate out the constraint $\delta(1-\sum_i x_i)$ 
in a way which preserves the property that all the endpoint singularities are located 
at the origin of parameter space, rather than creating singularities at $x_i=1$.

However, there are certain situations where it is useful to start the iterated decomposition 
at a later stage, where the delta-function already has been integrated out analytically,
or for an integrand which does not contain such a delta-function at all.
If one can find a convenient parametrisation and integrate out 
one Feynman parameter analy\-ti\-cally, this can be beneficial for complicated integrals, 
because it reduces the number of integration variables for the subsequent
Monte Carlo integration and therefore will improve the numerical efficiency. 

In such cases, the user can skip the primary sector decomposition step and 
insert the functions to be factorized directly into the Mathematica input file.
The purpose of this option is to be more flexible with regards to the 
functions to be integrated, such that expressions for loop integrals which are not in the ``standard form"
can be dealt with as well.
This includes the possibility to perform a deformation of the integration contour into the complex plane, 
taking the user-defined functions as a starting point.
Oriented at the functions $\mathcal{F}$ and $\mathcal{U}$ for 
the ``standard" loop case, 
the user-defined functions can encompass the product of two arbitrary 
polynomial functions with different exponents, and an additional numerator.

The command to launch the calculation is the same as for the standard setup, except that 
the Mathematica input file looks different (for an example see the file {\tt templateuserdefined.m}), 
and that the extension {\tt -u} should be appended:\\
{\it ./launch -p myparamfile.input -t mytemplateuserdefined.m -u} \\
The ``-u'' stands for ``user defined" and skips the primary sector decomposition step. 

\subsection{Scanning over ranges of numerical parameters}

The algebraic part of \secdec{} can deal with symbolic expressions for 
the kinematic invariants or other parameters contained in the integrand. Therefore, 
the decomposition and subtraction parts only need to be done once 
and for all, and then can be employed for  the calculation
of many different numerical points. 
The program comes with scripts which facilitate the scanning over ranges of numerical values 
for the kinematic invariants. 
There is a perl script {\tt helpmulti.pl} which can be used to produce an input file 
{\tt multiparam.input} containing ranges of numerical values for the invariants/symbolic parameters 
contained in the integrand.
This way the user does not have to type in all the numerical values by hand.
Each line in {\tt multiparam.input} defines a new run. 
An example of a {\tt multiparam.input} file is contained in both the {\tt loop} 
and the {\tt general} directories. 
A detailed description of this option is also given in \cite{Carter:2010hi}.

In order to launch the calculations for the values specified in {\tt multiparam.input}, 
the command \\
{\it perl multinumerics.pl [-d myworkingdir] -p multiparamfile} \\
should be issued.

Please note that before executing the script {\tt multinumerics.pl}, 
the Mathematica-generated functions must already
be existent. The simplest way to do this is to make one run with
{\it exeflag=1} in the single-run parameter file. 

To collect the results:
in single-machine mode ({\it clusterflag=0}), the results will be collated automatically 
and written to the {\tt graph} directory specified in the parameter file.
In {\it cluster} mode 
the {\it multinumerics.pl} script has to be run again with the argument `1' appended i.e.
{\it perl multinumerics.pl [-d myworkingdir] -p multiparamfile 1} to collect the results. 
In both cases files with the extension {\tt [pointname][i].gpdat} will be created for each 
coefficient of $\eps^{i}$ in the Laurent expansion, which can be plotted easily.
For example, if the parameter $s$ has been scanned over, and $s$ is the first element 
in the list of invariants, setting {\tt xplot=1} in {\tt multiparam.input}
will define $s$ to be on the x-axis, 
and therefore $s$ will be printed into the first column of the {\tt .gpdat} file. 
Thus, the columns in these result files will be \\
{\tt (s) (real part)  (error\_real\_part) (imaginary part)  (error\_imaginary\_part) (timings)}.
Result files for each individual point (extension {\tt .res}) will also be created.

\section{Applications}

\subsection{Two-loop four-point integrals with two mass scales}
\label{sec:hh}

Here we give an example for the definition of a non-planar two-loop four-point function, 
containing both internal masses, and massive external legs with a different mass.

\begin{figure}[htb!]
\begin{center}
\includegraphics[width=0.35\textwidth]{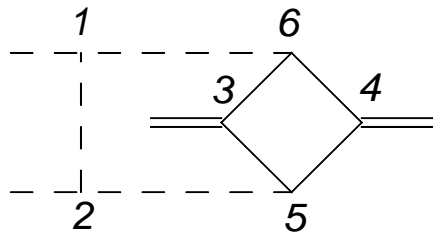}
\end{center}
\caption{Example of a non-planar two-loop box diagram with two mass scales. 
Dashed lines denote massless propagators, solid lines massive propagators. 
The double lines denote massive legs with a mass different from the internal mass.
	\label{fig:hhnp1} }
\end{figure}
\begin{figure}[htb!]
\includegraphics[width=0.38\textwidth,angle=-90]{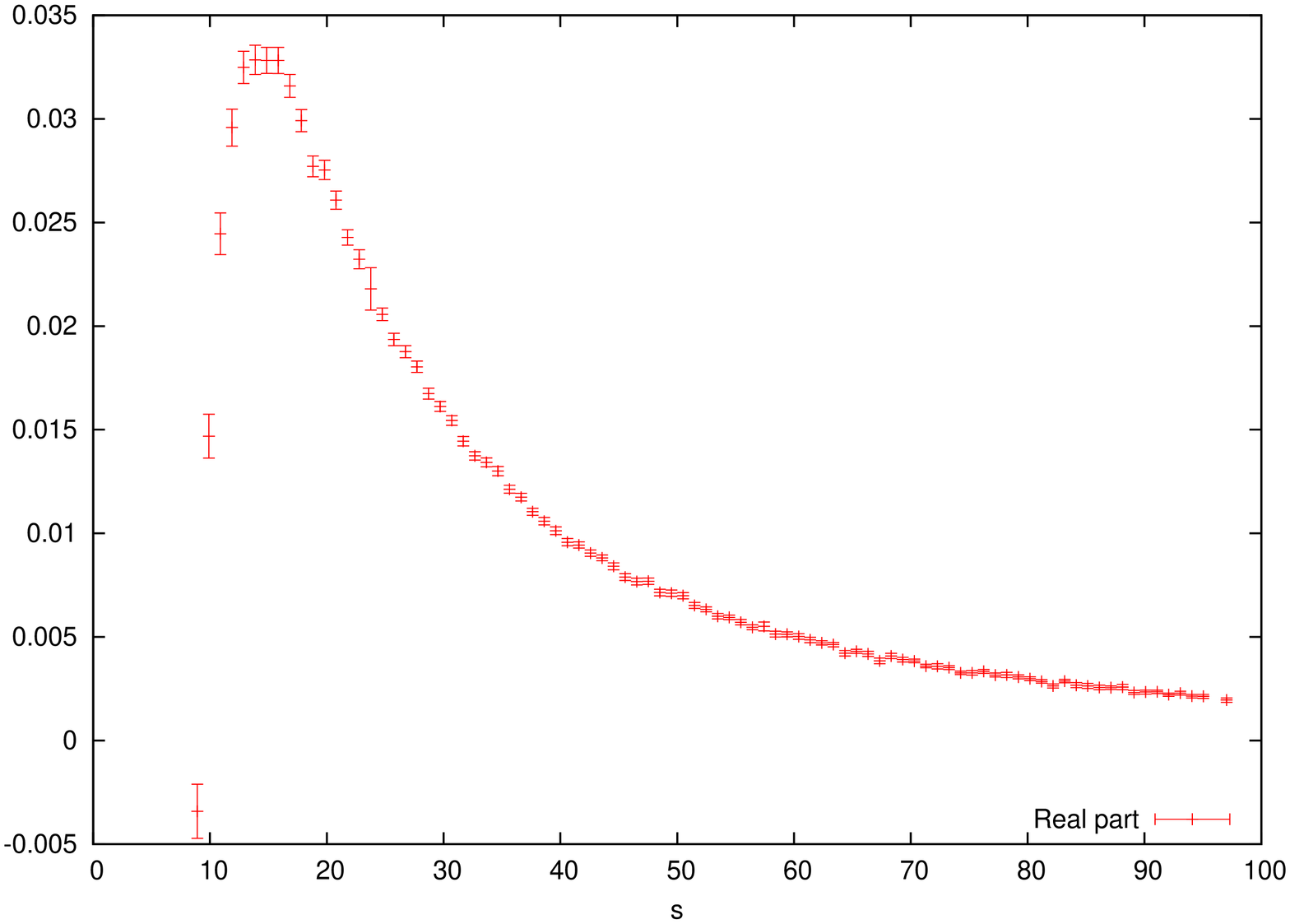}
\includegraphics[width=0.38\textwidth,angle=-90]{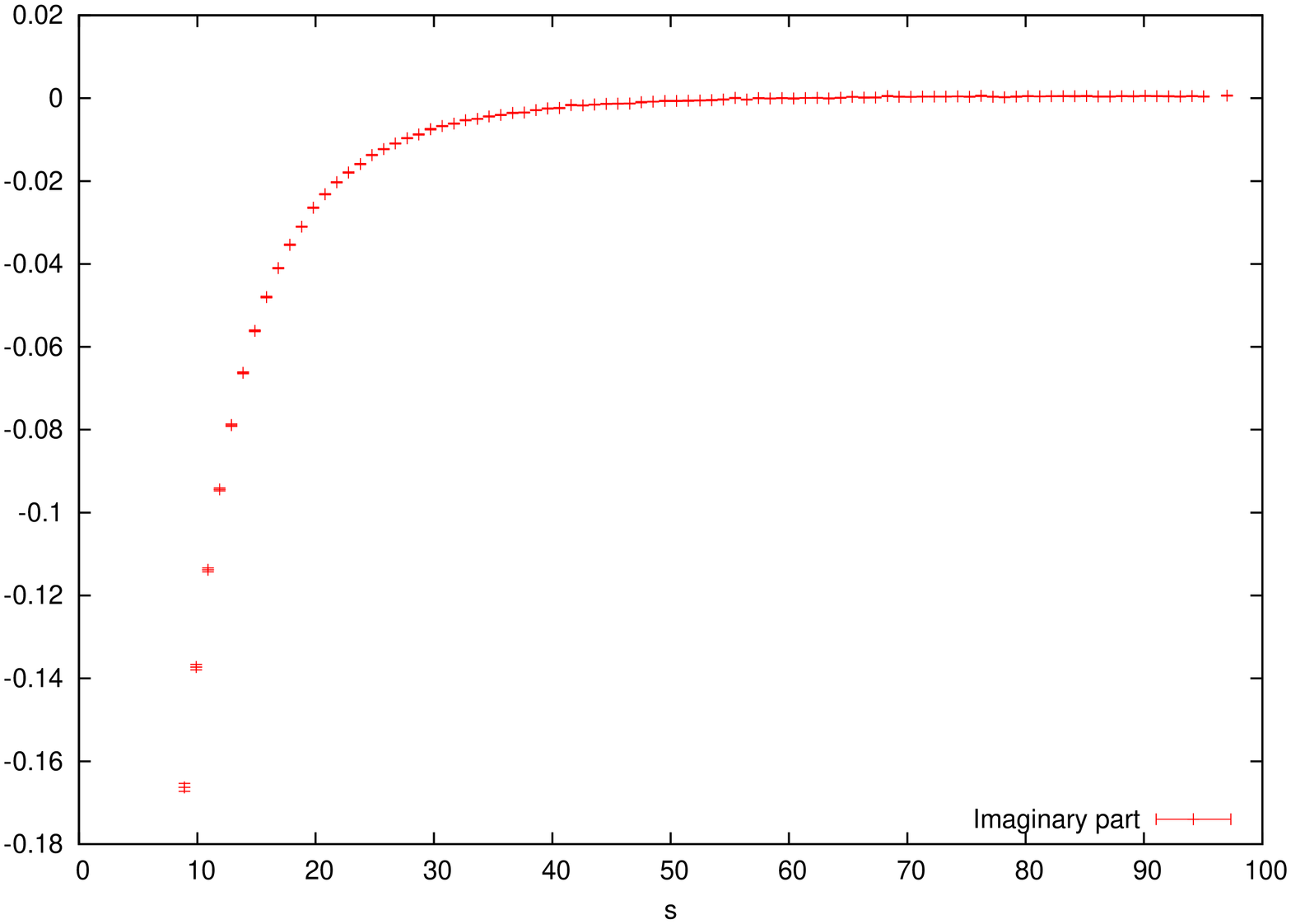}
\caption{Real and imaginary parts of the finite part of the graph shown in 
Fig.~\ref{fig:hhnp1}, for the values $m_1^2=1,m_2^2=0.522, t=-3.978, u=-s-t+2m_2^2$. \label{fig:resultshhnp1} }
\end{figure}
To define the scalar graph depicted in Fig.~\ref{fig:hhnp1}, the user can use the 
option {\tt cutconstruct=1} as explained above. To this end, the vertices 
containing external legs need to be labelled with the same number as the corresponding external leg. 
Then the vertices containing only internal lines are labelled (in arbitrary order). 
The graph is defined in {\tt mytemplate.m} by a list {\tt proplist} containing the two vertex labels 
$i_1,i_2$ a certain propagator is connecting, and the mass $m$ of that propagator, 
as $\{m,\{i_1,i_2\}\}$. For example, for the graph of Fig.~\ref{fig:hhnp1}
\begin{verbatim}
proplist={{0,{1,2}},{0,{1,6}},{0,{2,5}},{ms[1],{3,6}},{ms[1],{3,5}},
\end{verbatim}
\begin{verbatim}
          {ms[1],{4,5}},{ms[1],{4,6}}}; .
\end{verbatim} 
The kinematic conditions have to be given as 

\begin{verbatim}
onshell={ssp[1]->0,ssp[2]->0,ssp[3]->ms[2],ssp[4]->ms[2]}; .
\end{verbatim}
The expression {\tt ssp[i]} is the standard name for $p_i^2$. 
Squared masses $m_i^2$ are denoted by {\tt ms[i]}.
The restriction to ``standard names" will be lifted in the next release of \secdec{},
where the user will be able to define arbitrary symbols for the invariants.

\noindent Numerical results for this graph are shown in Fig.~\ref{fig:resultshhnp1}.

\subsection{Momentum dependent two-loop corrections to neutral Higgs boson masses in the MSSM}


In this section we briefly describe an application of \secdec{} where two-loop 
two-point integrals with up to four different mass scales are entering. 
Until recently, the \order{\alt\als} corrections to the neutral $\cp$-even Higgs boson masses in the MSSM
have been known adopting a full 
$\overline{\rm{DR}}$ scheme only \cite{Martin:2003it,Martin:2004kr,Martin:2005qm}, or neglecting the 
momentum dependence~\cite{Heinemeyer:1998np,Heinemeyer:1999be,Degrassi:2001yf}.
For a consistent inclusion of the corrections in the program 
\fh~\cite{Heinemeyer:1998np,Heinemeyer:1998yj,Hahn:2009zz,Hahn:2013ria,Degrassi:2002fi,Frank:2006yh}, 
the top-quarks and top-squarks need to be renormalized on-shell. These momentum dependent \order{\alt\als}
corrections missing so far in \fh{} 
were calculated in \cite{Borowka:2014wla}, where the 34 mass configurations of the analytically unknown two-loop 
two-point integrals are calculated numerically using \secdec.
For further details we refer to \cite{Borowka:2014wla,Borowka:2014aaa}.
Very recently, an independent calculation of these corrections appeared~\cite{Degrassi:2014pfa}, where the unknown
integrals have been calculated numerically using the program {\tt TSIL}~\cite{Martin:2005qm},
which is a dedicated program for the evaluation of two-loop two-point functions, 
based on differential equations. 

\vspace*{4mm}

At tree level, the mass matrix of the neutral $\cp$-even Higgs bosons in the $(\phi_1^0,\phi_2^0)$ basis can be written as 
\begin{align}
\label{eq:nondiag}
M_{\text{Higgs}}^{2,\text{tree}}=\left( \begin{matrix} M_A^2\text{sin}^2\, \beta + M_Z^2\text{cos}^2\, \beta & 
-(M_A^2 + M_Z^2)\,\text{sin}\, \beta \text{cos}\, \beta \\ 
-(M_A^2 + M_Z^2)\,\text{sin}\, \beta \text{cos}\, \beta & 
M_A^2\text{cos}^2\, \beta + M_Z^2\text{sin}^2\, \beta \end{matrix} \right)\;,
\end{align}
where $M_A$ is the mass of the $\cp$-odd neutral Higgs boson $A$. 
The rotation to the basis formed by the mass eigenstates $H^0,h^0$ is given by 
\begin{align}
\label{eq:physbasis}
 \left( \begin{matrix} H^0 \\ h^0 \end{matrix} \right) =  \left( \begin{matrix} \textrm{cos} \,\alpha  & 
 \textrm{sin} \,\alpha \\ 
-\textrm{sin} \,\alpha & 
\textrm{cos} \, \alpha \end{matrix} \right) \left( \begin{matrix} \phi_1^0 \\ 
\phi_2^0 \end{matrix} \right)\;.
\end{align}
The higher-order corrected $\cp$-even Higgs boson masses in the
MSSM are obtained  from the corresponding propagators
dressed by their self-energies. 
The inverse propagator matrix in the $(\Pe, \Pz)$ basis is given 
by
\begin{align}
\label{eq:prop}
(\Delta_{\text{Higgs}})^{-1} = -\text{i}
\left( \begin{matrix} 
p^2 - m_{\phi_1}^2 + \hat{\Sigma}_{\phi_1}(p^2) & -m_{\phi_1\phi_2}^2 +\hat{\Sigma}_{\phi_1\phi_2}(p^2)\\ 
-m_{\phi_1\phi_2}^2 +\hat{\Sigma}_{\phi_1\phi_2}(p^2) & p^2 - m_{\phi_2}^2 + \hat{\Sigma}_{\phi_2}(p^2) 
\end{matrix} \right) \text{ ,}
\end{align}
where the $\hat{\Sigma}(p^2)$ denote the momentum-dependent renormalized Higgs-boson 
self-energies, $p$ being the external momentum. The latter have been calculated at the two-loop level, 
at order $\als\alt$.

\begin{figure}[htb!]
\begin{center}
\subfigure[]{\raisebox{0pt}{\includegraphics[width=0.2\textwidth]{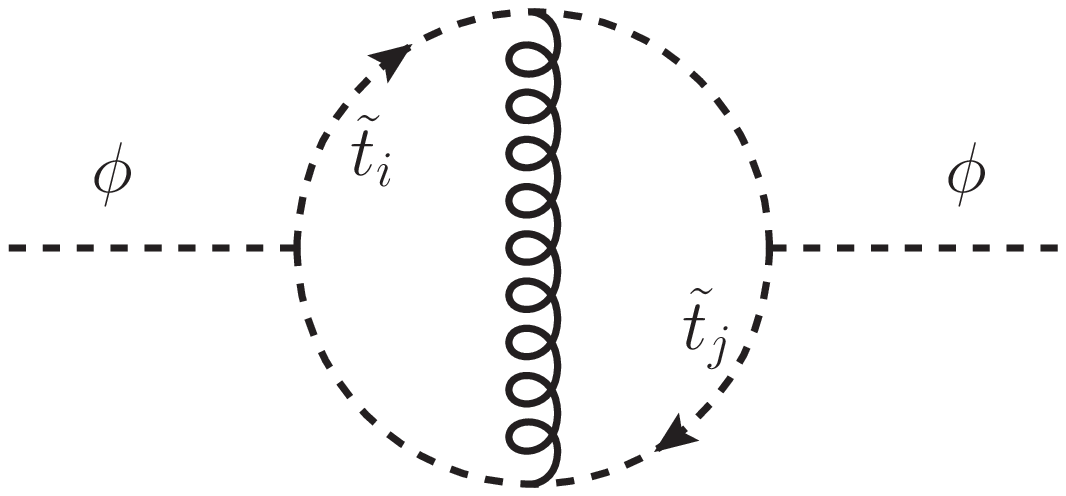}} }
\subfigure[]{\raisebox{1pt}{\includegraphics[width=0.2\textwidth]{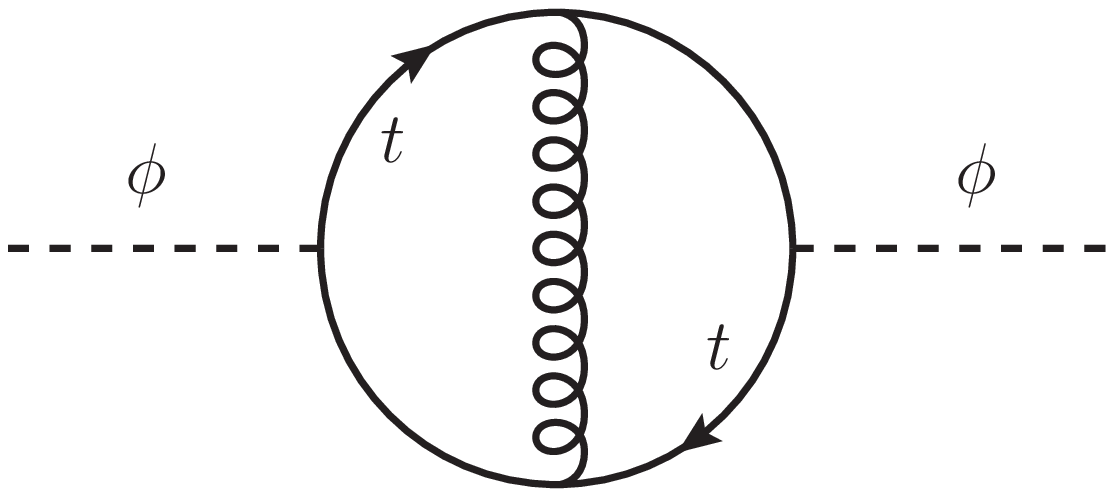}} }
\subfigure[]{\raisebox{1pt}{\includegraphics[width=0.2\textwidth]{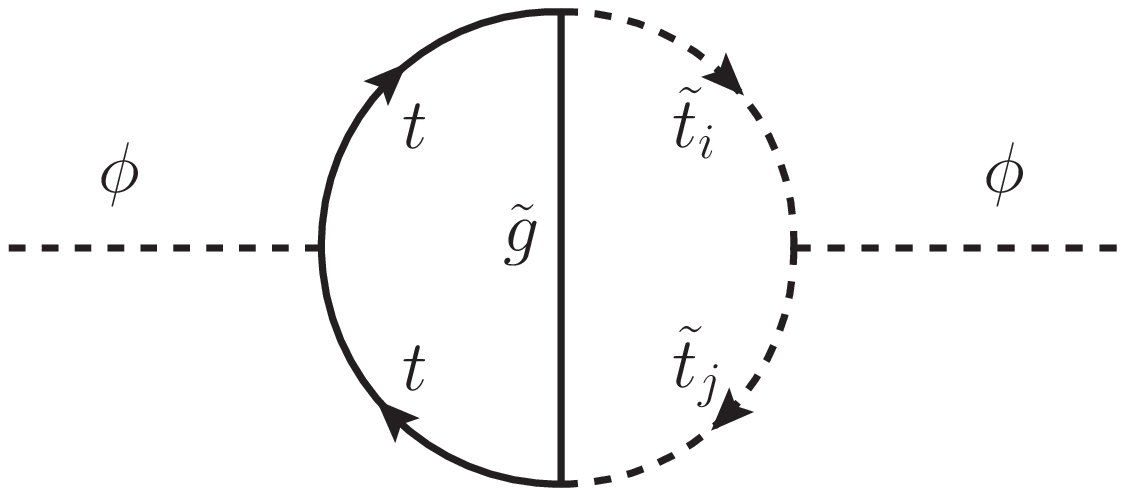}} }
\subfigure[]{\raisebox{-1pt}{\includegraphics[width=0.2\textwidth]{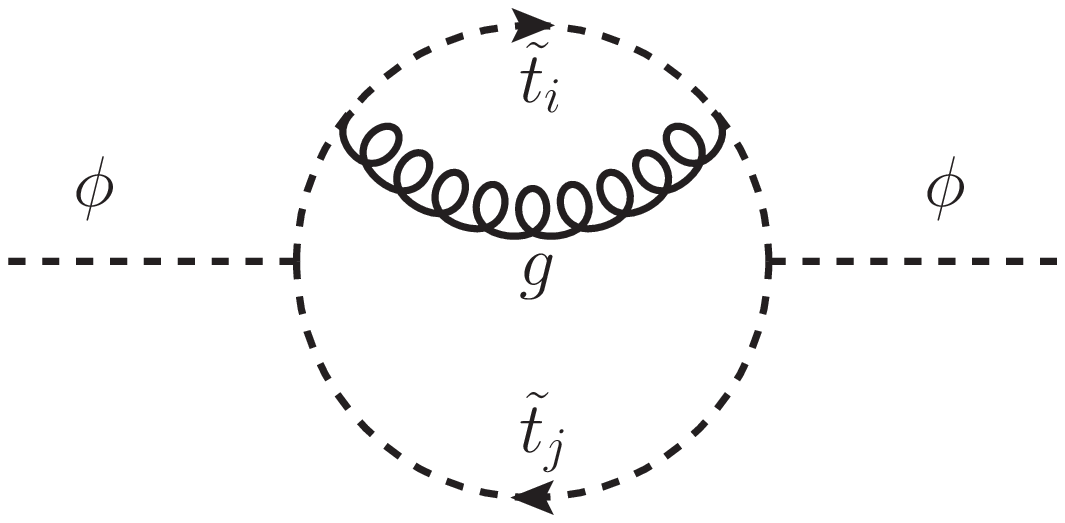}} }\\
\hspace{-10pt}\subfigure[]{\raisebox{0pt}{\includegraphics[width=0.2\textwidth]{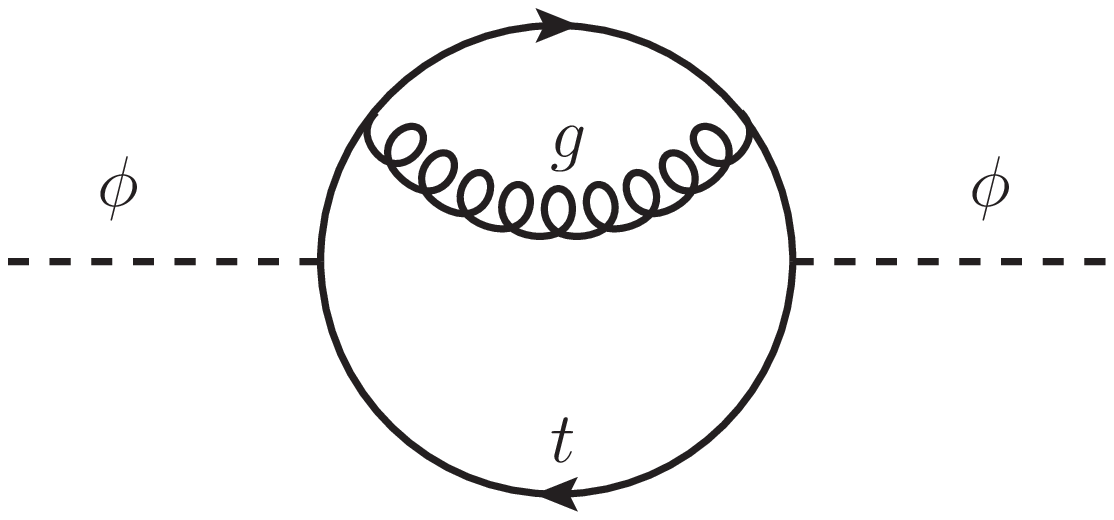}} }
\subfigure[]{\raisebox{-1pt}{\includegraphics[width=0.2\textwidth]{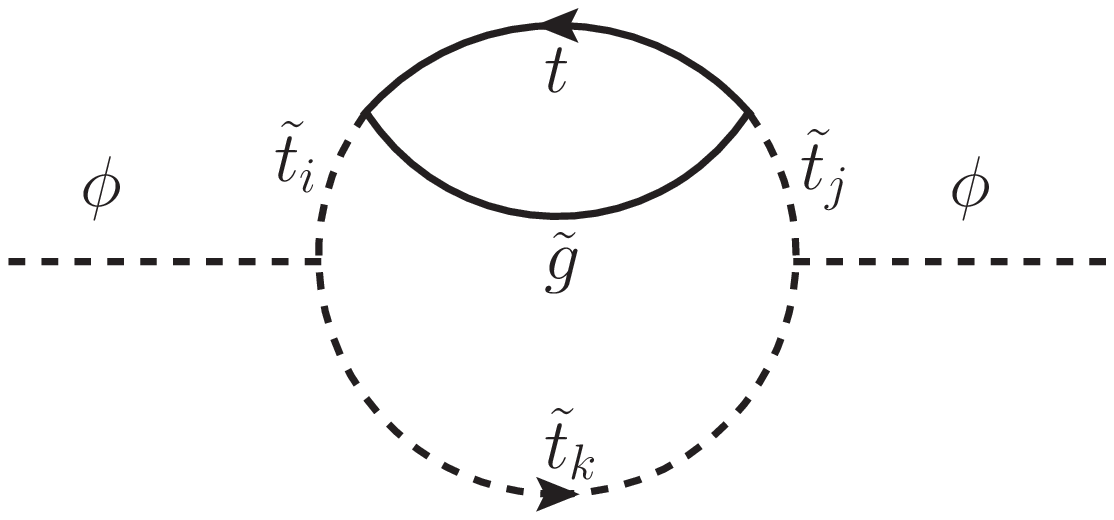}} }
\subfigure[]{\raisebox{0pt}{\includegraphics[width=0.2\textwidth]{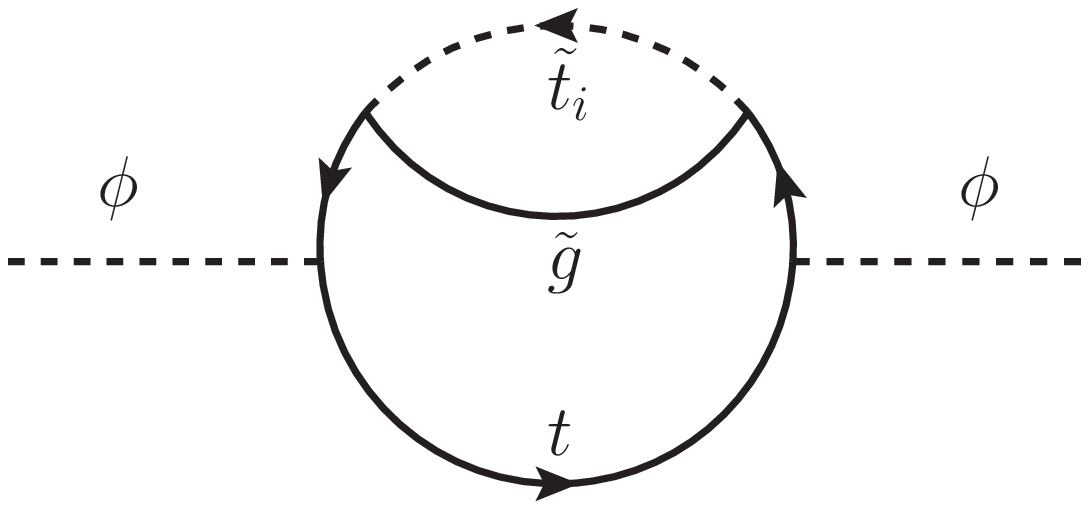}} }
\hspace{9pt}
\subfigure[]{\raisebox{0pt}{\includegraphics[width=0.15\textwidth]{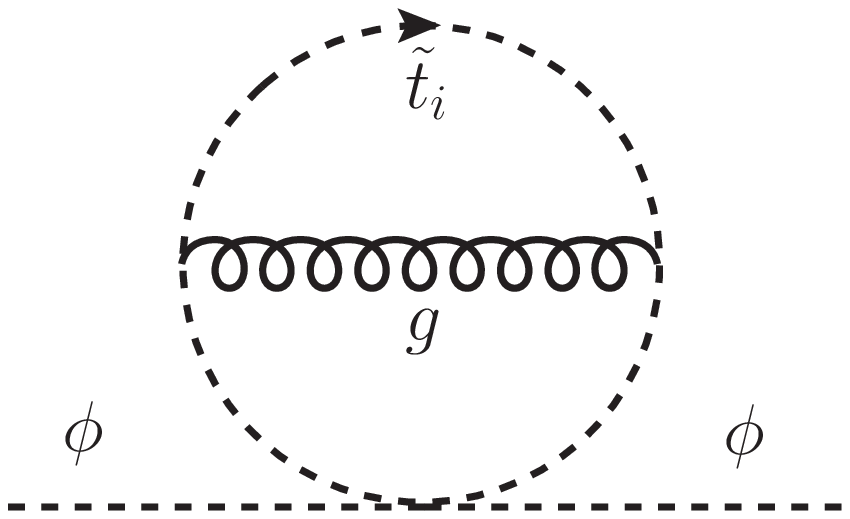}} }
\end{center} \vspace{-10pt}
\caption{Examples of  \twol\ diagrams enetring the Higgs-boson self-energies
($\phi = h, H, A$). }
\label{fig:fd_hHA}
\end{figure}

\begin{figure}[htb!]
\begin{center}
\begin{minipage}{0.99\textwidth}
\includegraphics[width=0.23\textwidth]{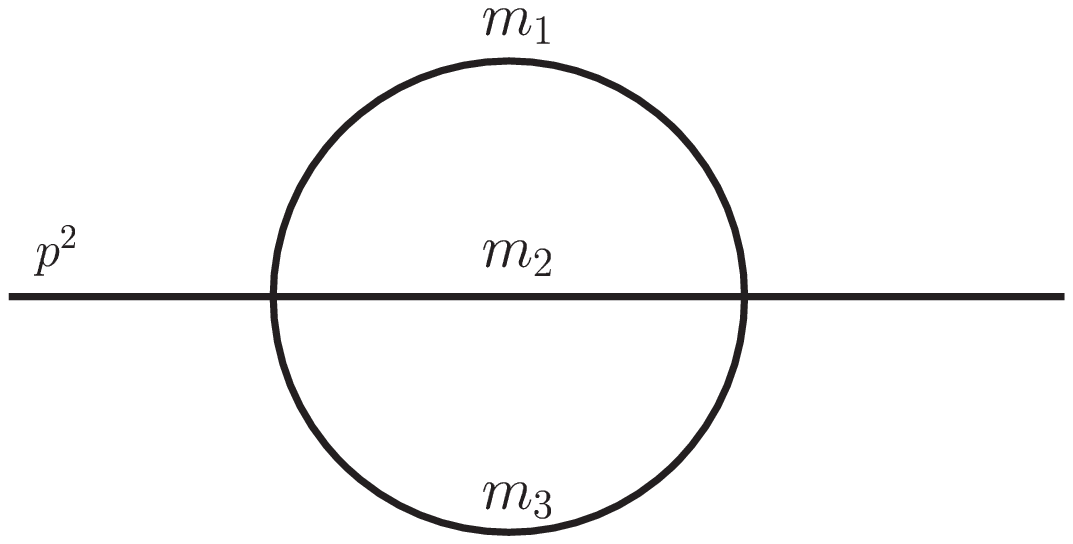}\hspace{1pt}
\includegraphics[width=0.23\textwidth]{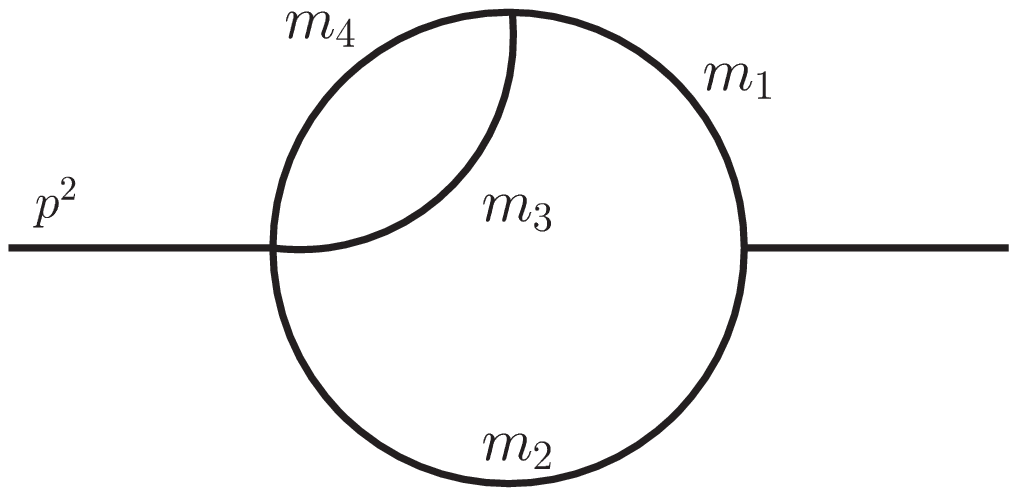}\hspace{1pt}
\includegraphics[width=0.23\textwidth]{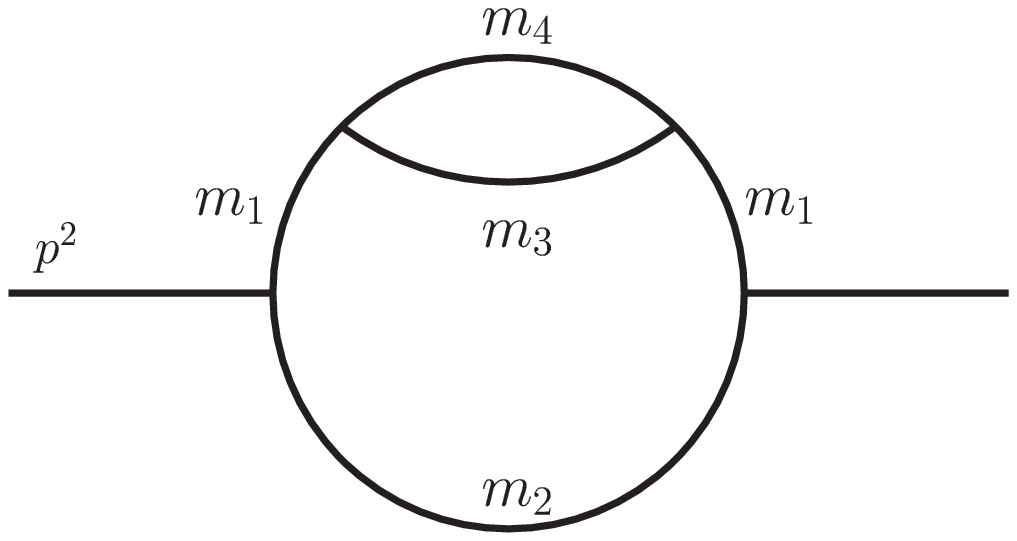}\hspace{1pt}
\includegraphics[width=0.23\textwidth]{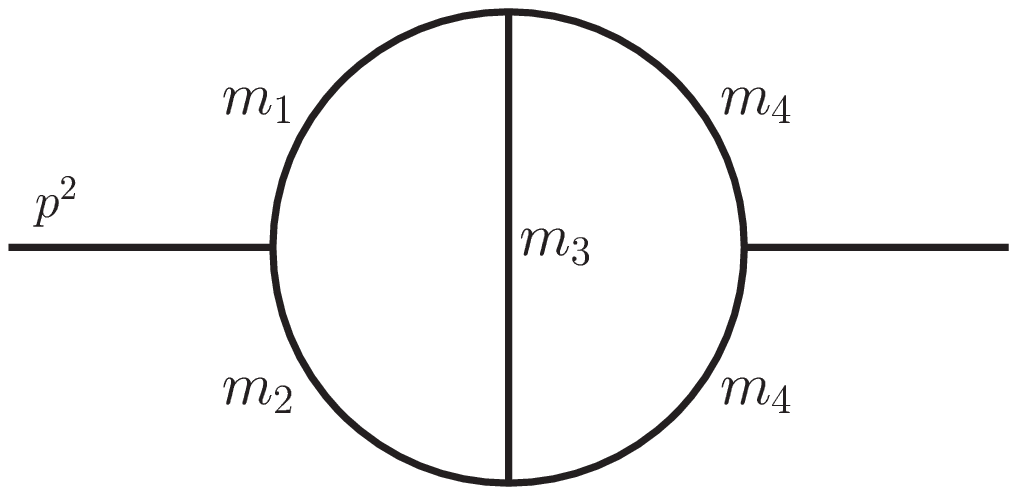}\vspace{-15pt}\\
\begin{center} 
\hspace*{-15pt}$T_{234}$ \hspace{85pt} $T_{1234}$\hspace{85pt} $T_{11234}$\hspace{80pt} $T_{12345}$
\vspace{10pt}\\
\end{center} 
\end{minipage}
\caption{Topologies which have been calculated numerically using \secdec.}
\label{fig:Tintegrals}
\end{center}
\end{figure}

Our calculation is performed in the Feynman-diagrammatic approach. 
We adopt a hybrid on-shell/$\overline{\rm{DR}}$ scheme, in line with the renormalization 
of previous higher-order contributions included in the program 
\fh\footnote{We neglect a numerically insignificant shift in the
value of $\tb$.}, 
see Ref.~\cite{Borowka:2014wla} for more details. 
To obtain expressions for the unrenormalized self-energies and tadpoles
at \order{\alt\als}, the evaluation of genuine two-loop diagrams
and one-loop graphs with counterterm insertions is required.
Example diagrams for the neutral Higgs-boson self-energies are shown
in Fig.~\ref{fig:fd_hHA}.
\newpage
For the counterterm-insertions, 
one-loop diagrams with external top quarks/squarks have 
to be evaluated as well.
The complete set of contributing Feynman diagrams 
has been generated with the
program {\tt FeynArts}~\cite{Hahn:2000kx,Hahn:2001rv} (using the model file including
counterterms from \citere{Fritzsche:2013fta}).
Tensor reduction and the evaluation of traces was done with 
support from the programs~{\tt FormCalc}~\cite{Hahn:1998yk} and 
{\tt TwoCalc}~\cite{Weiglein:1993hd}. 
The resulting two-loop integrals which depend on the external momentum
contain four topologies for which only partial analytical results are available.
These topologies are shown in Fig.~\ref{fig:Tintegrals}.
They occur in 34 different mass configurations, 
and have been evaluated with \secdec. 
 \vspace{-30pt}\\
\subsection*{Computation of mass shifts}
\label{subsec:howmassshifts}
The calculation of the self-energies is performed in the $(\phi_1^0,\phi_2^0)$ basis. 
To be consistent with all other higher-order contributions to the 
Higgs-boson masses incorporated in the  program \fh, 
the renormalized self-energies in the $(\phi_1^0,\phi_2^0)$ basis is
rotated into the physical $(h^0,H^0)$ basis, where the tree-level 
propagator matrix is diagonal, via
\begin{subequations}
\BEA
\ser{H^0H^0}^{(2)}&=& 
\cos^2\!\alpha \,\ser{\Pe\Pe}^{(2)} + 
\sin^2\!\alpha \,\ser{\Pz\Pz}^{(2)} + 
\sin (2 \alpha) \, \ser{\PePz}^{(2)} \text{ ,}\\
\ser{h^0h^0}^{(2)} &=& 
\sin^2\!\alpha \,\ser{\Pe\Pe}^{(2)} + 
\cos^2\!\alpha \,\ser{\Pz\Pz}^{(2)} - 
\sin (2 \alpha) \,   \ser{\PePz}^{(2)} \text{ ,} \\
\ser{h^0H^0}^{(2)} &=& 
\sin \alpha \cos\alpha \,(\ser{\Pz\Pz}^{(2)} - \ser{\Pe\Pe}^{(2)}) + 
\cos (2 \alpha) \,   \,\ser{\PePz}^{(2)} \text{ ,}
\EEA
\label{eq:transformationphi12tohH}%
\end{subequations}%
and $\alpha$ is the tree-level mixing angle. 

The resulting new contributions to the neutral $\cp$-even Higgs-boson 
self-energies, containing all momentum-dependent and additional constant 
terms, are assigned to the differences
\begin{equation}
\De\ser{ab}^{(2)}(p^2) = \ser{ab}^{(2)}(p^2) - \tilde\Sigma_{ab}^{(2)}(0)\,,
\qquad
ab = \{H^0H^0,h^0H^0,h^0h^0\}\,.
\label{eq:DeltaSE}
\end{equation}
Note the tilde (not hat) on $\tilde\Sigma^{(2)}(0)$, which signifies that 
not only the self-energies are evaluated at zero external momentum but
also the corresponding counter-terms,
following Refs.~\cite{Heinemeyer:1998jw,Heinemeyer:1998kz,
Heinemeyer:1998np}.
A finite shift $\De\hat{\Sigma}^{(2)}  (0)$
therefore remains in the limit $p^2\to 0$ 
due to $\de m_{A^0}^{2(2)} = \re\se{A^0A^0}^{(2)}(m_{A^0}^2)$ being computed 
at $p^2=m_{A^0}^2$ 
in $\hat\Sigma^{(2)}$, but at $p^2=0$ in $\tilde\Sigma^{(2)}$.

\medskip

According to Eq.~(\ref{eq:prop}), 
the $\cp$-even Higgs boson masses are determined by the
poles of the $h^0$-$H^0$-propagator matrix. 
This is equivalent to solving the equation
\begin{equation}
\left[p^2 - m_{h^0}^2 + \hSi_{h^0h^0}(p^2) \right]
\left[p^2 - m_{H^0}^2 + \hSi_{H^0H^0}(p^2) \right] -
\left[\hSi_{h^0H^0}(p^2)\right]^2 = 0\,~,
\label{eq:proppole}
\end{equation}
yielding the loop-corrected pole masses, $\Mh$ and $\MH$.
%

\subsection*{Numerical results}
\label{sec:numericalresults}

In our numerical analysis 
we find that  the effects on the light
$\cp$-even Higgs boson mass, $\Mh$, are sensitive to the value of the
gluino mass, $\Mgl$. For values of $\Mgl \sim 1.5 \tev$ corrections to
$\Mh$ of about $ -50 \mev$ are found, at the level of the anticipated
future ILC accuracy. For very large gluino masses, $\Mgl \gtrsim 4 \tev$, 
on the other hand, substantially larger corrections are found, at the
level of the current experimental accuracy of $\sim 500 \mev$. 
The new results of \order{\alt\als} including momentum dependence 
have been implemented into the program \fh.

\section{Conclusions}
We have described the program \secdec{}, with emphasis on useful features and new applications.
\secdec{} is a flexible tool which can be used to factorise poles from dimensionally regulated
parameter integrals, and to evaluate the pole coefficients numerically. 
As there is no restriction on the kinematic invariants, the program is particularly
useful for two-loop integrals with several mass scales, where analytic approaches 
have difficulties.

The program further contains scripts to facilitate scans over large ranges of numerical parameters, 
based on functions for the Laurent series expansion of the 
integral which can be produced once and for all.

The forthcoming version 3.0 of the program will contain further improvements, 
in particular offer new decomposition strategies.

\ack{
We would like to thank the organizers of ACAT2014 for the nice conference. 
We also thank Thomas Hahn, Sven Heinemeyer and Wolfgang Hollik 
for collaboration on the MSSM Higgs project, and Johannes Schlenk for 
his contributions towards \secdec{} version 3.0.
}

\section*{References}

\providecommand{\newblock}{}

\end{document}

%% file: definitions.tex
\newcommand{\bq}{\begin{align}}
\newcommand{\eq}{\end{align}}

\catcode`@=11
\def\citer{\@ifnextchar [{\@tempswatrue\@citexr}{\@tempswafalse\@citexr[]}}
 
\def\@citexr[#1]#2{\if@filesw\immediate\write\@auxout{\string\citation{#2}}\fi
  \def\@citea{}\@cite{\@for\@citeb:=#2\do
    {\@citea\def\@citea{--\penalty\@m}\@ifundefined
       {b@\@citeb}{{\bf ?}\@warning
       {Citation `\@citeb' on page \thepage \space undefined}}%
\hbox{\csname b@\@citeb\endcsname}}}{#1}}
\catcode`@=12

\def\citere#1{\mbox{Ref.~\cite{#1}}}

\newcommand{\Pe}{\phi_1^0}
\newcommand{\Pz}{\phi_2^0}

\newcommand{\PePz}{\phi_1^0\phi_2^0}

\def\order#1{\ensuremath{{\cal O}(#1)}}
\newcommand{\cp}{{\cal CP}}

\newcommand{\twol}{two-loop}

\newcommand{\fh}{{\sc FeynHiggs}}

\newcommand{\Mh}{M_h}
\newcommand{\MH}{M_H}

\newcommand{\Mgl}{m_{\tilde{g}}}

\newcommand{\tsf}{\theta\kern-.20em_{\tilde{f}}}
\newcommand{\tsfp}{\theta\kern-.20em_{\tilde{f}\prime}}
\newcommand{\tsq}{\theta\kern-.15em_{\tilde{q}}}

\newcommand{\VL}{\left( \begin{array}{c}}
\newcommand{\VR}{\end{array} \right)}
\newcommand{\ML}{\left( \begin{array}{cc}}
\newcommand{\MLd}{\left( \begin{array}{ccc}}
\newcommand{\MLv}{\left( \begin{array}{cccc}}
\newcommand{\MR}{\end{array} \right)}
\newcommand{\re}{\mathop{\mathrm{Re}}}

\newcommand{\tb}{\tan \beta}

\newcommand{\tev}{\,\, \mathrm{TeV}}

\newcommand{\mev}{\,\, \mathrm{MeV}}

\newcommand{\BC}{\begin{center}}
\newcommand{\EC}{\end{center}}
\newcommand{\BE}{\begin{equation}}
\newcommand{\BEA}{\begin{eqnarray}}
\newcommand{\EEA}{\end{eqnarray}}

\newcommand{\id}{{\rm 1\kern-.12em
\rule{0.3pt}{1.5ex}\raisebox{0.0ex}{\rule{0.1em}{0.3pt}}}}

\def\als{\alpha_s}
\def\alt{\alpha_t}

\def\de{\delta}

\def\eps{\varepsilon}

\def\De{\Delta}

\def\hSi{\hat{\Sigma}}
\newcommand{\se}[1]{\Sigma_{#1}}

\newcommand{\ser}[1]{\hat{\Sigma}_{#1}}